\documentclass[superscriptaddress,twocolumn,aps,prb,ltxgrid,showpacs]{revtex4}
\usepackage{epsfig}
\usepackage{amsbsy,amssymb,amsmath,bm,ulem}

\usepackage{color}

\newcommand{\sign}{\ensuremath{\mbox{\text{sign}}}}

\newcommand{\matr}[9]{\ensuremath{\left(\begin{array}{ccc} {#1} & {#2} & {#3} \\ {#4} & {#5} & {#6} \\{#7} & {#8} & {#9} \end{array}\right)}}
\newcommand{\vectr}[3]{\ensuremath{\left(\begin{array}{ccc} {#1} \\ {#2} \\ {#3}  \end{array}\right)}}
\renewcommand{\Re}{\mathop{\mathrm{Re}}}
\renewcommand{\Im}{\mathop{\mathrm{Im}}}

\graphicspath{{pic/}}

\begin{document}

\title{Nonlinearly driven Landau-Zener transition with telegraph noise}
\author{J. I. Vestg{\aa}rden}
\affiliation{Department of Physics and Center for Advanced
Materials and Nanotechnology,    University of Oslo, P. O. Box
1048 Blindern, 0316 Oslo, Norway}
\author{J. Bergli}
\affiliation{Department of Physics and Center for Advanced
Materials and Nanotechnology,    University of Oslo, P. O. Box
1048 Blindern, 0316 Oslo, Norway}
\author{Y. M. Galperin}
\affiliation{Department of Physics and Center for Advanced
Materials and Nanotechnology,    University of Oslo, P. O. Box
1048 Blindern, 0316 Oslo, Norway}
\affiliation{A. F. Ioffe
Physico-Technical Institute of Russian Academy of Sciences, 194021
St. Petersburg, Russia} \affiliation{Argonne National Laboratory,
9700 S. Cass Av., Argonne, IL 60439, USA}

\begin{abstract}
We study Landau-Zener like dynamics of a qubit influenced by
transverse random telegraph noise.  The telegraph noise is
characterized by its coupling strength, $v$ and switching rate,
$\gamma$. The qubit energy levels are driven nonlinearly in time,
$\propto \sign(t)|t|^\nu$, and we derive the transition probability in the
limit of sufficiently fast noise, for arbitrary exponent $\nu$. 
The longitudinal coherence after transition depends strongly on $\nu$
, and there exists a critical $\nu_c$ with qualitative difference between $\nu< \nu_c$
and $\nu > \nu_c$.  When $\nu<\nu_c$ the end state is always fully
incoherent with equal population of both quantum levels, even for 
arbitrarily weak noise. For $\nu>\nu_c$ the system keeps some coherence 
depending on the strength of the noise, 
and in the limit of weak noise no transition takes place. 
For fast noise $\nu_c=1/2$, while for slow noise $\nu_c<1/2$ and it depends on $\gamma$.
We also discuss transverse coherence, which is relevant when 
the qubit has a nonzero minimum energy gap. The qualitative dependency 
on $\nu$ is the same for transverse as for longitudinal coherence. 
The state after transition does in general depend on $\gamma$.
For fixed $v$, increasing $\gamma$ decreases the final
state coherence when $\nu<1$ and increase the final state coherence
when $\nu>1$. Only the conventional linear driving is independent of $\gamma$.
\end{abstract}

\pacs{03.65.Yz,85.25.Cp,05.40.Ca}

\maketitle

\section{Introduction}

Driven quantum systems are exceedingly more complicated to study than
stationary systems, and only few such problems 
have been solved exactly. An important exception
is the Landau-Zener transitions.\cite{zener32,landau32-ii,stueckelberg32}
In the conventional Landau-Zener problem, a two-level system is 
driven by changing an external parameter in such a way that the 
level separation $\Delta$ is a linear function of time, $\Delta(t)=at$. 
Close to the crossing point of the two levels an inter-level tunneling 
matrix element $g$ lifts the degeneracy in an avoided level crossing.
When the system is initially in the ground state the probability to find it 
in the excited state after the transition is 
$\exp(-\pi g^2/2a)$. Hence, fast rate drives the system to the excited state,
while the system ends in ground state when driven slowly.
The Landau-Zener formalism was originally 
developed for molecular and atomic physics, 
but has since then been applied to various systems and
many generalizations of the linearly driven 
two-level system exists, like avoided level 
crossing of multiple levels,\cite{brundobler93,shytov04}
repeated crossings,\cite{shytov03}
non-linear model, \cite{liu02}
and non-linear driving functions.\cite{garanin02}

In connection with decoherence of qubits there has recently been
increased interest in Landau-Zener transitions in systems coupled to
an environment. This problem is both of theoretical interest and of
practical importance for qubit experiments.\cite{sillanpaa06} The
noise affects the qubit in two ways.  First, it destroys coherence by
random additions to the phase difference of the two states
(dephasing). Second, it causes transitions and alters the level
occupation (relaxation).  The noisy Landau-Zener problem has been
discussed by several authors\cite{kayanuma84,kayanuma85,shimshoni91,shimshoni93,wubs06}
both for quantum and classical environments. In this work we will 
study classical noise processes. In particular, we will use a random 
telegraph process as the noise source. This allows us to study 
the effect of noise with long correlation time (slow, or non-Gaussian noise). 
In the limit of short correlation times we will recover the results of   
Pokrovsky and Sinitsyn\cite{pokrovsky03} who have considered this problem in the 
limit of fast noise.
An important result of their analysis was that there is a characteristic
time scale, $t_\text{noise}$, during which the noise is active.  If
this time scale is long compared to the time of the Landau-Zener
transitions, $t_\text{LZ}$, dynamics can be separated in a
noise-dominated regime for long times and a pure, noiseless
Landau-Zener transition for short times. This allows one to study separately
transitions driven purely by noise and the usual Landau-Zener transitions
driven by the tunneling amplitude $g$. We will follow this 
approach, which simplifies the problem considerably. 

Most works on noisy Landau-Zener transitions
are mainly concerned with transition probabilities. 
However, in the case of an open system 
it is also interesting to study the amount of decoherence, 
or purity, of the state after the transition is passed. 
In terms of the Bloch vector, the transition
probability is given by the $z$-component of the vector whereas the 
purity is given by its length. By generalization from the stationary case,
it is clear that longitudinal noise 
(noise in the level spacing $\Delta$) will cause dephasing at all times, 
and the final state will always be on the axis of the Bloch sphere, i.e., 
the $x$- and $y$-components of the Bloch vector decay to zero. 
For transverse noise (noise in the anticrossing energy $g$) 
the situation is less evident since the effect of the noise 
is reduced by the factor $g/\Delta$. 
When $\Delta$ increases sufficiently fast as function of time one can in a 
sense `run away' from the noise, and the final state will not decohere 
maximally. This motivates us to study the effect of nonlinear 
time dependences for the level splitting, similar to those considered in 
Ref.~\onlinecite{garanin02} for Landau-Zener transitions without noise. 
In particular, we will study power-law 
driving functions, $\Delta\sim \sign (t) |t|^\nu$, 
and we will find that there exists a critical  
$\nu_c$ such that the system is completely 
decohered for $\nu<\nu_c$ even for arbitrarily weak noise coupling.
For $\nu>\nu_c$ some coherence is retained. 
The critical $\nu_c$ will depend on the 
correlation time of the noise.

\section{Model}

\subsection{Hamiltonian}
Consider a solid state qubit, e.g., a Josephson charge qubit.\cite{sillanpaa06,faoro05,shnirman02}
The qubit is modeled as a two-level system and it couples to environment through
a randomly fluctuating addition $\chi(t)$ on its off-diagonal terms.
Let us here only consider dynamics for one realization of $\chi(t)$, while
in next section we will use the particular model
of random telegraph noise to derive master equations 
for the noise averaged quantities. 

The Hamiltonian is
\begin{equation}  \label{hamiltonian}
  H = \frac{1}{2}\Delta_\nu(t)~\sigma_z+\frac{1}{2}[g+\chi(t)]~\sigma_x
\end{equation}
where $\sigma_x$ and $\sigma_z$ are Pauli matrices,
$\Delta_\nu(t)$ is the diagonal splitting, 
$g$ is the minimal energy gap at the avoided level crossing.

The interesting dynamics comes from a power-law 
time-dependency 
\begin{equation}  \label{delta}
  \Delta_\nu(t) = \alpha_\nu |\alpha_\nu t|^\nu \mbox{sign}(t)
\end{equation}
with sweep rate $\alpha_\nu$ and exponent $\nu$. 
Linear sweep and no noise give exactly the Landau-Zener dynamics.
However, our focus will be on entirely noise-driven transition
for any exponent.

From here and throughout this work the quantum state is described by the
Bloch vector $\mathbf{r}\equiv (x,y,z)$. The Bloch vector is is given from 
the density matrix $\rho$ as 
\begin{equation}
\begin{split}
  x &= 2\Re\rho_{12} 
  , \\
  y &= 2\Im\rho_{12} 
  , \\
  z &= \rho_{11}-\rho_{22}
  .
\end{split}
\end{equation}
For a pure quantum system the vector {\bf r} is a unit vector. Under 
the influence of  noise its average value 
is in general less than unity. 

The dynamics of {\bf r} is given by the Bloch equation,
\begin{equation}  \label{bloch-equation}
   \dot {\mathbf r} = -\mathbf r \times \mathbf B
  ,
\end{equation}
analogous to a spin precessing in 
magnetic field $\mathbf B(t) = (g+\chi(t),0,\Delta_\nu(t))$.
We use units where $\hbar = 1$ throughout this work.

\subsection{Telegraph noise}

The noise model applied in this work is {\it random telegraph noise}.
Such noise occurs when defects create 
bistable traps, atomic or electronic,  in solids,
and is assumed\cite{harlingen04} to be a basic
source for various kinds of high and low-frequency noise.\cite{kogan}
For example, a large number of fast fluctuators 
with a narrow distribution of switching rates give 
Gaussian white noise. A broad distributions of switching rates can, on
the contrary, give rise to non-Gaussian, $1/f$ noise.\cite{paladino02,galperin06}
In experiments on solid state qubits,
the low-frequency $1/f$ noise is often the dominant
source of decoherence.\cite{nakamura02} For tiny devices, a small number, 
or even single fluctuators, can be important.
Relevant for transverse noise on Josephson charge qubits
telegraph noise characteristics has been
measured for electrons trapped in Josephson junctions,\cite{wakai86}
for intrinsic Josephson junctions in granular high-$T_c$ superconductors,\cite{jung96}
and for trapped single flux quanta.\cite{johnson90} 

If the bistable system, or \textit{fluctuator}, is more strongly
coupled to its surroundings than to the qubit we can consider its
dynamics to be independent of the qubit, and it will act as a
classical noise source, driven by its environment.  With this
approximation, the effect of the fluctuator on the qubit appears
through a randomly switching addition $\pm v$ to the tunneling
energy. The constant, $v$, represents fluctuator-qubit coupling strength,
which will be called noise strength for short.  We assume the
switchings between the two fluctuator states to be independent, random
events. The rates of random switching is assumed to be the same
between both fluctuator levels, $\gamma_{+-}=\gamma_{-+}=\gamma$.
This holds when the fluctuator level-spacing is small compared to
the temperature.  Our fluctuator model is thus a stochastic process and
the probability $P_k$ to switch $k$ times in a time interval $t$ is
given by the Poisson distribution,
\begin{equation}\label{poisson}
  P_k=\frac{(\gamma t)^k}{k!}e^{-\gamma t} \,.
\end{equation}
The telegraph process has the property
$\chi(t)\chi(0)=\pm v^2$, where the $+$ and $-$ sign
are for a even and odd number of switches, respectively.
Hence, the {\it autocorrelator} is
\begin{equation}
  \begin{split}
    S(t) 
    &=\langle \chi(t)\chi(0)\rangle 
    = \sum_{k=0}^\infty \chi_k(t)\chi_k(0)P_k 
    ,  \\
    &= v^2~\sum_{k=0}^\infty (-1)^k\frac{(\gamma t)^k}{k!}e^{-\gamma t} 
    = v^2e^{-2\gamma t}
    ,
   \label{autocorrelator}
\end{split}
\end{equation}
for $t>0$.
Correspondingly, the cosine transform of Eq.~\eqref{autocorrelator} 
(the noise power spectrum) is a Lorentzian
\begin{equation}
  \hat S(\omega)=\int_0^{\infty} \! dt \, S(t)\cos(\omega t) =
  v^2\frac{2\gamma}{(2\gamma)^2+\omega^2}\,   .
  \label{spectral-density}
\end{equation}
The noise power spectrum is important since all results for fast noise can 
be expressed by this function. 

It must be noted that for many qubit experiments the environment cannot 
be considered as classical and a quantum description of noise is necessary.\cite{astafiev04} 
The Spin-Boson model was discussed in Ref.~\onlinecite{shnirman02} for stationary system and in 
Ref.~\onlinecite{wubs06} in connection with Landau-Zener transitions.
Ref.~\onlinecite{grishin05} has developed a model for fluctuating charges at 
finite temperature. Random telegraph noise is the high
temperature limit of this model.

\subsection{Master equations}

We will now average Eq.~\eqref{bloch-equation} over the noise
and derive master equations for a qubit 
coupled to one random telegraph process.
The quantum state is now only known with 
a certain probability and we need to operate 
with averaged quantities rather than the pure quantum states.
The average value of $\mathbf r$ is 
\begin{equation}
  \mathbf r_p = \langle\mathbf r\rangle = \int d^3r~p(\mathbf r,t)~\mathbf r
  .
\end{equation}
where $p=p(\mathbf r,t)$ is the probability of being in Bloch state {\bf r} at time $t$.

For the particular model of one random telegraph process
there are two possible values of the effective magnetic field acting upon
the  qubit according to Eq. (\ref{bloch-equation}):
\begin{equation}
  \mathbf B_\pm=\mathbf B_0 \pm \mathbf v \, .
\end{equation}
where $\mathbf v$ is a constant vector.
Here $\mathbf B_0(t)=(g,0,\Delta_\nu(t))$ controls 
the time evolution of the quantum mechanical system.
We will now derive the set of master equations. The derivation is in fact
valid for any two-level system coupled to one fluctuator in 
arbitrary direction, not just Landau-Zener like dynamics 
and transverse noise. The derivation follows
Refs.~\onlinecite{bergli06} and \onlinecite{bergli07}.

Let $p=p(\mathbf r,t)$ be the probability to be in
$\mathbf r$ at time $t$. 
Now split $p(\mathbf r,t)=p_+(\mathbf r,t)+p_-(\mathbf r,t)$
where $p_+(\mathbf r,t)$ and $p_-(\mathbf r,t)$ are the probabilities 
to be in state $\mathbf r$ at time $t$ under 
rotation around $\mathbf B_+$ and $\mathbf B_-$,
respectively.

The master equations for $p_+$ and $p_-$ are 
\[
\begin{split}
  p_+(\mathbf r,t+\epsilon) &= \alpha p_+(\mathbf r-\delta \mathbf r_+,t) 
  + \beta p_-(\mathbf r-\delta \mathbf r_-,t) 
  ,\\
  p_-(\mathbf r,t+\epsilon) &= \alpha p_-(\mathbf r-\delta \mathbf r_-,t) 
  + \beta p_+(\mathbf r-\delta \mathbf r_+,t)  
  ,
\end{split}
\]
where $\epsilon$ is a small time change and $\alpha$ and $\beta$ are the 
staying and switching probabilities, respectively.
When $\epsilon\ll\gamma$ we can neglect multiple switchings, and 
Eq.~(\ref{poisson}) can be expanded to give 
$\alpha \approx P_0\approx 1-\gamma \epsilon$ and $\beta \approx P_1\approx
 \gamma \epsilon$.
The spatial changes $\delta \mathbf r_\pm$ represent the vector's displacements 
during the time interval $\epsilon$. This  
is given from the Bloch equation, Eq.~\eqref{bloch-equation}, as 
$ \delta \mathbf r_\pm = -\mathbf r\times \mathbf B_\pm \epsilon$. 
Expanding to first order in $\epsilon$ gives
\[
\begin{split}
  \dot p_+ &= -\gamma p_+ + \gamma p_- + (\mathbf r\times\mathbf B_+)\cdot \nabla p_+ 
  ,\\
  \dot p_- &= -\gamma p_- + \gamma p_+ + (\mathbf r\times\mathbf B_-)\cdot \nabla p_-
  .
\end{split}
\]

The probabilities enable us to define equations for the
averaged quantities $\mathbf r_\pm=\int d^3r~\mathbf rp_\pm$,  
\begin{equation}
\begin{split}
   \dot {\mathbf r}_+ &= -\gamma  \mathbf r_+ + \gamma \mathbf  r_- - (\mathbf r_+\times\mathbf B_+)
  ,\\
   \dot {\mathbf r}_- &= -\gamma  \mathbf r_- + \gamma \mathbf r_+ - (\mathbf r_-\times\mathbf B_-)
  \label{eq-r_+-r_-}
  .
\end{split}
\end{equation}
The quantities $\mathbf r_+$ and $\mathbf r_-$ are just auxiliary quantities 
and the final master equations are expressed by $\mathbf r_p=\mathbf r_++\mathbf r_-$ 
and $\mathbf r_q=\mathbf r_+-\mathbf r_-$.
The quantities normally measured in 
experiment are those quantities averaged over $p$, and $\mathbf r_p$ are the 
averaged components of the Bloch vector. 
Isolating $\mathbf r_p$ and $\mathbf r_q$ yields 
\begin{equation}
\begin{split}
   \dot {\mathbf r}_p &= -\mathbf r_p\times \mathbf B_0 - \mathbf r_q\times \mathbf v 
  ,\\
   \dot {\mathbf r}_q &= -2\gamma\mathbf r_q -  \mathbf r_q\times\mathbf B_0 - \mathbf r_p\times \mathbf v
  \label{eq-r_p-r_q}
  .
\end{split}
\end{equation}
The above equations are exact for one telegraph process.
Compared to the noiseless case, the number of equations rise from two 
(i.e., three equation and constraint of $|\mathbf r|=1$) to
six equations. Adding more fluctuators, the number of equations 
will grow exponentially.\cite{bergli06}
\begin{figure}[t]
  \centering
  \epsfig{file=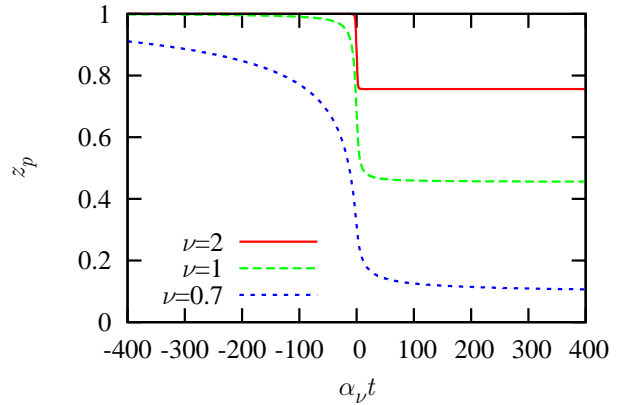, width = 8.5cm}
  \caption{\label{fig:fast-of-t}
    The $z_p(t)$ as a function of time for fast noise, 
    $\gamma/\alpha_\nu = 2$ and $v/\alpha_\nu = 0.5$.
    The transition time extends significantly with decreasing $\nu$.
  }  
\end{figure}

\subsection{Master equations for simplified problem}

Let us now study the simplified problem
of entirely noise-driven transition, i.e., 
$g=0$. In this case the set of
six equations, Eq.~\eqref{eq-r_p-r_q},
decouple in two sets of equations in
$(x_p,y_p,z_q)$ and $(x_q,y_q,z_p)$, respectively.
A system initially prepared in one 
energy eigenstate has $x_p=y_p=0$. Assuming also the initial state 
of the fluctuator to be random we have 
$z_q(-\infty)=0$, which means that $x_p$ and $y_p$ 
remain zero as long as $g=0$. Thus coherence only relays on $z_p$ and
we will for the following concentrate on the set $(x_q,y_q,z_p)$. 
The master equations are
\begin{equation}
  \vectr{\dot x_q}{\dot y_q}{\dot z_p} 
  =
  \matr{-2\gamma}{-\Delta_\nu}{0}{\Delta_\nu}{-2\gamma}{-v}{0}{v}{0}\vectr{x_q}{y_q}{z_p}\, .
  \label{x_q-y_q-z_p}
\end{equation}
Isolating $z_p$ yields the integral equation
\begin{equation}
  \begin{split}
    \dot z_p
    &= -\int_{-\infty}^tdt_1~\cos(\theta(t)-\theta(t_1) )~S(t-t_1)~z_p(t_1) \\
    &= -\int_0^{\infty}dt_2~\cos(\theta(t)-\theta(t-t_2) )~S(t_2)~z_p(t-t_2)
    ,
  \end{split}
  \label{integral-z_p}
\end{equation}
where
\begin{equation} \label{deftheta}
  \theta(t)=\int_0^tdt'\Delta_\nu(t')=\frac{1}{\nu+1}|\alpha_\nu t|^{\nu+1}
\end{equation}
and $S(t)$ given by Eq.~\eqref{autocorrelator}.
The integral equation, Eq.~\eqref{integral-z_p}, is exact for one
telegraph process, 
and valid for all transition rates. The equation
is the same as found in Ref.~\onlinecite{pokrovsky03} for any
fast noise source. Hence, all conclusions drawn from Eq.~\eqref{integral-z_p} 
in the limit $\gamma\to\infty$ are also valid for any Gaussian noise source. 
\begin{figure}[t]
  \centering
  \epsfig{file=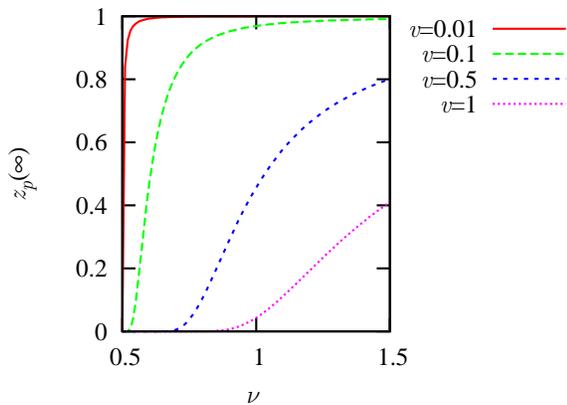, width = 8cm}
  \caption{\label{fig:fast-of-nu-by-v}
    The $z_p(\infty)$, Eq.~\eqref{z_p-infty}, as function of $\nu$, for fast noise; 
    $\gamma/\alpha_\nu = 10$. The weak noise has a strong $\nu$-dependency near $\nu=1/2$.
  }  
\end{figure}

\section{Fast noise}

With fast noise we mean finite but large $\gamma$, 
$\gamma \gg \alpha_\nu$.  
Then the relevant contributions in the integral of Eq.~\eqref{integral-z_p} 
are for small $t_2$. Series expansions in $t_2$ yields
\begin{equation} \label{z_p-general}
  \frac{\dot z_p}{z_p}
  \approx - \int_0^{\infty}dt_2~\cos(\Delta_\nu(t) t_2)~S(t_2) = -\hat S(\Delta_\nu(t))
  .
\end{equation}
The solution is
\begin{equation} \label{z_pt}
  z_p(t)=\exp\left[-\int_{-\infty}^tdt'~\hat S(\Delta_\nu(t')) \right]
  ,
\end{equation}
with noise power spectrum $\hat S$. 
Recalling\cite{shnirman02} that the relaxation rate 
of a qubit without driving is $\Gamma_\text{relax}=\hat S(E)$ at the qubit level spacing $E$ we 
can understand the above expression as the total relaxation over many short 
time intervals, the relaxation rate in each interval being given by the 
usual expression for the static case. This can only be done in the limit of 
fast noise. 

For the particular model of random telegraph noise
$\hat S$ is given by Eq.~\eqref{spectral-density}
and Eq.~\eqref{z_p-general} reads as
\begin{equation} \label{zpdot2}
  \frac{\dot z_p}{ z_p }
  = -v^2 \frac{2\gamma}{(2\gamma)^2+\Delta_\nu^2(t)}
  .
\end{equation}
The full integrated Eq.~\eqref{zpdot2} is expressed through
hypergeometric functions, which will not be written here.
A numerical solution is plotted 
in Fig.~\ref{fig:fast-of-t} for various exponents. 
It illustrates that the fast noise curves are smooth 
and all fluctuations are averaged out. Also, 
it shows that transitions times get longer for decreasing $\nu$.

The most interesting quantity, however, 
is the value at infinity which for $\nu>1/2$ is 
\begin{equation}
  z_p(\infty) =
  \exp\left[-2\frac{v^2}{\alpha_\nu^2}\left(\frac{2\gamma}{\alpha_\nu}\right)^{1/\nu-1} \! 
    \frac{\pi/2\nu}{\sin(\pi/2\nu)}\right]
  .
  \label{z_p-infty}
\end{equation}
This equation makes it possible to explore how the 
final state depends on $v$, $\gamma$, and $\nu$. 

For $\nu<1/2$ the integral of Eq. (\ref{zpdot2}) diverges 
and we get $z_p(\infty)=0$, independently of $v$ and $\gamma$.
When $z_p(\infty)=0$ both levels are occupied with 
same probability and this represents a fully incoherent state.
The fact that the result is independent of $v$ means 
that  arbitrarily weak noise destroys coherence completely.
This is similar to a stationary system where noise always dominates at long times.  
The result is actually a bit surprising. It is obvious that 
a static system finally looses all coherence. However, in this case
the energy levels split by up to square root
of time and even this is not enough to avoid total 
decoherence. For $\nu>1/2$ the results are no longer independent of 
$v$ and $\gamma$. In this sense one can say that the 
regimes for $\nu<1/2$ and $\nu > 1/2$ are qualitatively different.
Thus we identify the critical $\nu_c=1/2$ in the limit of fast noise. 

Fig.~\ref{fig:fast-of-nu-by-v} shows $z_p(\infty)$ as a function of $\nu$. 
For decreasing $v$ the change near $\nu=1/2$ get sharper and 
in the limit $v\to 0$ it approaches a step function of $\nu$.

Another interesting feature of  Eq.~\eqref{z_p-infty} 
is how $z_p(\infty)$ changes with increasing $\gamma$. 
For $\nu<1$ increasing $\gamma$ means that $z_p(\infty)$ decreases 
and goes to zero in the extremely fast noise limit, $\gamma/\alpha_\nu\to\infty$.
In other words, faster noise reduces end state coherence. The opposite is the 
case for $\nu>1$. Then faster noise increases the end state coherence 
and in fact $z_p(\infty)\to 1$ when  $\gamma/\alpha_\nu\to\infty$.
This behavior is to some extent counterintuitive since one could initially 
expect faster noise would always decrease coherence.
The linear driving is truly a special case since $z_p(\infty)$ is 
independent of $\gamma$ for $\nu=1$.
Note that in Ref.~\onlinecite{pokrovsky03} where the case $\nu=1$ was 
considered, the limit $\gamma\rightarrow\infty$ was taken together 
with the limit $v\rightarrow\infty$ in such a way that $v^2/\gamma$
remained constant. In their case, $z_p(\infty)$ depends on $\gamma$ 
and goes to 0 when  $\gamma\rightarrow\infty$.

From the denominator of Eq.~\eqref{zpdot2} one can identify 
a time scale characteristic for the action of the noise, 
$t_\text{noise}=\alpha_\nu^{-1}(2\gamma/\alpha_\nu)^{1/\nu}$. Thus
$t_\text{noise}$ increases with increasing $\gamma$ and 
decreasing $\nu$. For very large times, $t\gg t_\text{noise}$, the 
$z(t)$ will approach its end value as power of time.  
Integration of Eq.~\eqref{zpdot2} in this limit yields 
the asymptotic solution 
\begin{equation}
  z_p(t)=z_p(\infty)\left(1+\frac{1}{2\nu-1}\frac{2\gamma}{\alpha_\nu}
    \left(\frac{v}{\alpha_\nu}\right)^2(\alpha_\nu t)^{1-2\nu}\right)
  \label{z_p-long-times}
  ,
\end{equation}
with $z_p(\infty)$ given by Eq.~\eqref{z_p-infty}.
Eq.~\eqref{z_p-long-times} illustrates again the message of Fig.~\ref{fig:fast-of-t},
namely that convergence gets slower for decreasing $\nu$ and near 
the critical value of $\nu=1/2$ the transition is very slow.
For $\nu<1/2$, the expansion,  Eq.~\eqref{z_p-long-times},
is not valid. 

For the important linear case there is also a
nice explicit solution of Eq.~\eqref{zpdot2} for all times,
\begin{equation}
  z_p(t)= 
  \exp\left(-\frac{v^2}{\alpha_1^2}\left[\frac{\pi}{2} 
      + \arctan\left(\frac{\alpha_1^2}{2\gamma}~t\right)\right]\right)
  ,
\end{equation}
in which the end state simplifies to
\begin{equation}
  z_p(\infty) = e^{-\pi\left(v/\alpha_1\right)^2}
  .
\end{equation}

\begin{figure}[t]
  \centering
  \epsfig{file=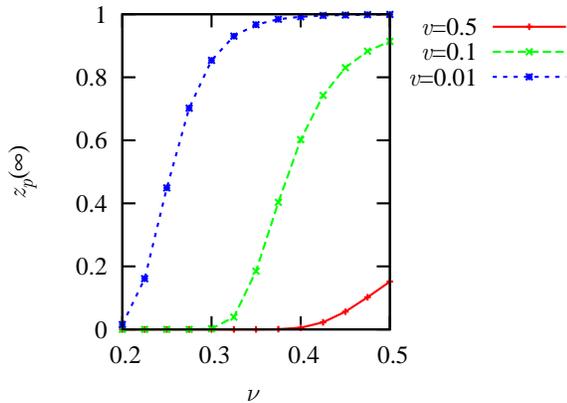, width = 8cm}
  \caption{\label{fig:slow-of-nu-by-v}
    The $z_p(\infty)$ as function of $\nu$ for weak and slow noise; $\gamma/\alpha_\nu = 0.1$.
    Obtained by numerical integration of Eq.~\eqref{integral-z_p}. The plot shows 
    a critical value of $\nu_c\approx 0.2$ which is less than the value for fast 
    noise seen in Fig.~\ref{fig:fast-of-nu-by-v}.
  }  
\end{figure}

\section{Slow and weak noise}

Now we will study the influence of one slowly varying telegraph process,
$\gamma \lesssim \alpha_\nu $ 
in the limit of weak noise, $v\ll \alpha_\nu$. 
We start with Eq.~\eqref{integral-z_p}, 
which  is exact for both fast and slow telegraph noise. 
A series expansion in $v/\alpha_\nu$ yields
\begin{equation} \label{smallv}
  z_p(\infty)
  \approx 1-v^2\int_{-\infty}^{\infty} \! \! dt \int_{-\infty}^t \! \! dt_1
  \cos[\theta(t)-\theta(t_1) ]\, e^{-2\gamma(t-t_1)}
  ,
\end{equation}
with $\theta$ defined in Eq.~\eqref{deftheta}.

In the extreme limit $\gamma = 0$ the equations are the same 
as for the nonlinear Landau-Zener system without noise. 
In this limit the integral Eq.~(\ref{smallv}) can be solved exactly, recovering 
the results of Ref.~\onlinecite{garanin02}:
\begin{equation} \label{z_p-weak-slow}
  z_p(\infty)=1-2\left(\frac{v}{\alpha_\nu}\right)^2
  \left[(1+\nu)^{-\frac{\nu}{\nu+1}}~\Gamma\left(\frac{1}{\nu+1}\right)\right]^2
  .
\end{equation}
where $\Gamma$ is the gamma function. Eq.~\eqref{z_p-weak-slow}
shows only weak $\nu$-dependency. Thus the $\nu$-dependency for a finite and 
small $\gamma$ will also be weak. The reason is that 
the first order in $\gamma$ will also be proportional to 
the a power of the small factor  $(v/\alpha_\nu)$.
The expression Eq.~\eqref{z_p-weak-slow} is only 
approximately valid for small, but finite, $\gamma$, provided 
that $\nu>\nu_c$.

Let $\gamma$ be small but nonzero. As for fast noise we define the
critical $\nu_c$ by $z_p(\infty)=0$ for all $\nu<\nu_c$
independently of $v$. Hence, $\nu_c$ can be identified by studying the
convergence of Eq.~\eqref{smallv}. The integral diverges for
$\nu<\nu_c$ and converges for
$\nu>\nu_c$.\cite{note:lz-slow-weak-convergence} We have not been able
to analyze the convergence of this integral analytically.  Instead,
Eq.~\eqref{integral-z_p} is solved numerically for a selected small
value of $\gamma$. This value gives a hint of how $\nu_c$ depends on
$\gamma$. Practically, $\nu_c$ is found by plotting $z_p(\infty)$ as a
function of $\nu$ for fixed $\gamma/\alpha_\nu$ and decreasing values
of $v/\alpha_\nu$.  The plot in Fig.~\ref{fig:slow-of-nu-by-v} shows
the expected behavior: $z_p(\infty)$ decreases when $\nu$ decreases,
and goes to zero at finite $\nu$, even for very small values of
$v/\alpha_\nu$.  The behavior is analogous to the fast noise plot of
Fig.~\ref{fig:fast-of-nu-by-v}, but the critical value is
significantly lower. For $\gamma/\alpha_\nu=0.1$ we find $\nu_c\approx
0.2$. This lowering is expected since $\nu_c=0$ for $\gamma=0$.  It
must be noted that there is large numerical inaccuracy for the low
$\nu$ in Fig.~\ref{fig:slow-of-nu-by-v}, since the integral is close
to divergency.

\begin{figure}[t]
  \centering
  \epsfig{file=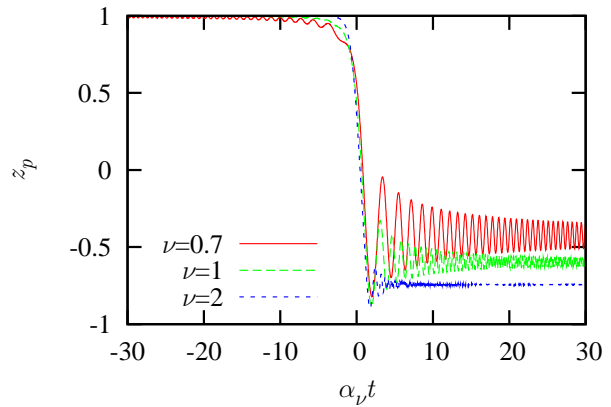, width = 8.5cm}
  \caption{The $z_p(t)$ as a function of time for slow and strong noise; 
    $\gamma=0$ and $v/\alpha_\nu=1$.
    This case is mathematically equivalent to a Landau-Zener transition
    and rapid oscillations are observed, unlike for fast noise, cf. Fig.~\ref{fig:fast-of-t}. 
  }
  \label{fig:slow-of-t}
\end{figure}

\section{Slow and strong noise}

\begin{figure}[t]
  \centering
  \epsfig{file=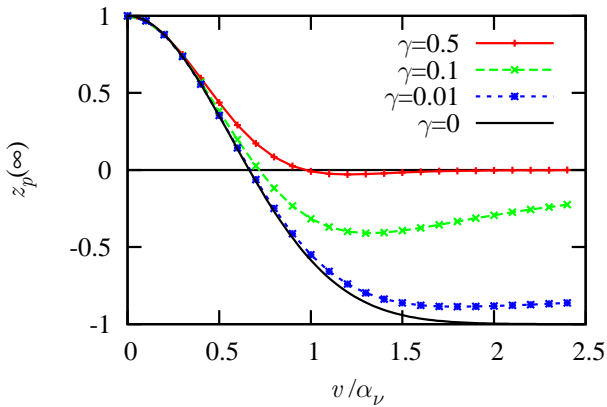, width = 8.5cm}
  \caption{The $z_p(\infty)$ as function of $v/\alpha_\nu$, for slow noise; 
    $\gamma/\alpha_\nu=0.1$ and $\nu=1$. Obtained by numerical integration
    of Eq.~\eqref{integral-z_p}.
    For  $\gamma/\alpha_\nu\ll 1$  
    the value $z_p(\infty)$ can take any value between -1 and 1, not just those 
    in the upper half of the Bloch sphere.    
  }
  \label{fig:slow-of-v-by-gamma}
\end{figure}

Let us again look at slow noise, $\gamma\lesssim \alpha_\nu$,
but without restrictions on $v/\alpha_\nu$.  
In particular we are interested in the regime in which $v$ is of 
same order of magnitude as $\alpha_\nu$. In this regime 
the results depend strongly on the actual values 
of $\alpha_\nu$, $v$, $\gamma$, and $\nu$. 
The transitions are quite sharp and give rapid oscillations 
after the transition, as seen in Fig.~\ref{fig:slow-of-t},
contrary to the smoothened transitions of the fast noise, 
exemplified in Fig.~\ref{fig:fast-of-t}.
Unlike fast noise the results depends strongly on $\gamma$
also for $\nu=1$. Fig.~\ref{fig:slow-of-v-by-gamma} shows
how $z_p(\infty)$ depends on $v/\alpha_\nu$ for slow noise
and linear driving. One first thing to notice is that slow noise, 
contrary to fast noise, can drive the system to the other diabatic level. 
This is seen as $z_p(\infty)<0$ in the plot. Second, some curves 
for $z_p(\infty)$ go through the center of the Bloch-sphere when $v$ increases.
The center of the Bloch sphere represents maximum decoherence since both states
are occupied with equal probability. Consequently, under some conditions 
increasing noise strength will also increase the system purity after transition.

\section{Transverse coherence}

We will now discuss transverse coherence (phase coherence). 
It is given by $r_\bot=\sqrt{x_p^2+y_p^2}$, where $x_p$ and $y_p$ are the transverse 
components of the Bloch vector. In particular we are interested 
in the behavior for fast noise and long times and see if we can identify a critical $\nu_c$,
as we did for the longitudinal coherence.

Transverse coherence becomes relevant when there is a nonzero anticrossing 
energy $g$ in the Hamiltonian Eq.~\eqref{hamiltonian}. 
In that case the Bloch vector makes a rotation away from the $z$-axis,
acquiring nonzero $r_\bot$.
This rotation is a Landau-Zener transition. 
A full solutions of the master equations, Eq.~\eqref{eq-r_p-r_q}, 
with $g\neq 0$ is difficult, and in the spirit of Pokrovsky and Sinitsyn \cite{pokrovsky03} 
we consider the case where the characteristic time $t_\text{LZ}$ of the Landau-Zener transition 
is much shorter than the time over which the noise is effective,  $t_\text{noise}$.
In principle, this would mean that we should study Eq.~\eqref{eq-r_p-r_q} in the case 
$g=0$, starting at $t=t_0$, where $t_\text{LZ}\ll t_0\ll t_\text{noise}$.
As long as we are only interested in determining
the critical $\nu_c$ and not in the precise value of the transition probability
we can therefore consider a Bloch vector 
starting in the equatorial plane of the Bloch sphere, 
$r_\bot(0)=r_0$ and $z_q(0)=0$ at time $t=0$. 
With this starting point we now assume $g=0$ and 
$\Delta_\nu$ again as an arbitrary power of time.
From Eq.~\eqref{eq-r_p-r_q} we have
\begin{equation}
  \vectr{\dot x_p}{\dot y_p}{\dot z_q} 
  =
  \matr{0}{-\Delta_\nu}{0}{\Delta_\nu}{0}{-v}{0}{v}{-2\gamma}\vectr{x_p}{y_p}{z_q}
  ,
  \label{x_p-y_p-z_q}
\end{equation}
which should be compared with Eq.~\eqref{x_q-y_q-z_p} for $(x_q,y_q,z_p)$.
Isolating $z_q$ gives
\begin{equation} \label{z_q-tmp}
  \begin{split}
  z_q(t)=v\int_0^{t}dt_2~y_p(t-t_2)e^{-2\gamma t_2}
  .
\end{split}
\end{equation}
For fast noise and long times the important contributions 
again come from small $t_2$. However, we 
must be careful when doing expansions of $y_p(t)$ since 
the product $\Delta_\nu (t)t_2$ is not necessarily small.
Let us define $A(t)=x_p(t)+iy_p(t)$ and explicitly take out the 
problematic, long-time phase factor $\theta(t)=\int_0^tdt'\Delta_\nu(t')$:  
\begin{equation}
  A(t)=r_\bot(t)~e^{i\theta(t)+i\varphi(t)}
  ,
\end{equation}
where $\varphi(t)$ is a phase factor that varies less rapidly 
than $\theta(t)$. Now expanding at long times $t\gg t_2$,
\begin{equation}
  A(t-t_2)\approx r_\bot(t)~e^{i\theta(t)+i\varphi(t)-i\Delta_\nu(t)t_2}
  .
\end{equation}
Inserting this into Eq.~\eqref{z_q-tmp} and isolating $r_\bot$ yields
\begin{equation} \label{r_bot0}
  \begin{split}
    \frac{\dot r_\bot}{r_\bot} 
    &= -\frac{v^2}{(2\gamma)^2+\Delta_\nu^2(t)}\sin(\theta+\varphi) \\
    &\times \left\{
      2\gamma\sin(\theta+\varphi)-\Delta_\nu(t)\cos(\theta+\varphi)
    \right\}
    .
  \end{split}
\end{equation}
At long times the sine and cosine functions oscillate rapidly and we 
substitute these terms with their respective average values, giving the final
equation for $r_\bot$,
\begin{equation}  \label{r_bot}
  \frac{\dot r_\bot}{r_\bot} 
  = -\frac{v^2}{2}\frac{2\gamma}{(2\gamma)^2+\Delta_\nu^2(t)}
  = -\frac{1}{2}\hat S(\Delta_\nu(t))
  .
\end{equation}
where $\hat S$ is the noise power spectrum. 

Eq.~\eqref{r_bot} has the same form as Eq.~\eqref{z_p-general} 
for $z_p$, so the whole discussion of Eq.~\eqref{z_p-general} is in fact valid 
also for Eq.~\eqref{r_bot}. In particular this means they share the same 
critical value. Thus $\nu_c=1/2$ for both transverse 
and longitudinal coherence; when $\nu<\nu_c=1/2$, the system end state 
is fully incoherent no matter the value of $v$ and $\gamma$.
We have not searched for the critical exponent of the transverse coherence 
for slow noise, $\gamma\lesssim\alpha_\nu$. However, if it exists it need not 
have the same numerical value as for the longitudinal coherence.

The right hand side of Eq.~\eqref{r_bot} can be interpreted as 
the instantaneous dephasing rate. In that case one recovers\cite{shnirman02} 
the result from transverse noise without driving, $\Gamma_\varphi = \hat S(E)/2$,
where $E$ is qubit level spacing.
The relation to the instantaneous relaxation rate is 
$\Gamma_\varphi=\Gamma_\text{relax}/2$; exactly the same as 
for the weak coupling limit of a Gaussian noise source.

There is one more thing to note about Eq.~\eqref{r_bot}.
The approximations needed to get to this expressions 
are coarser than those for $z_p$. In fact, the fast noise 
regime of $z_p$ start at $\alpha_\nu t\gtrsim 1$, while 
for $r_\bot$ it must be truly large, $\alpha_\nu t\gg 1$.

\section{Summary}

We have considered Landau-Zener like dynamics of a qubit in noisy
environment. The environment is modeled as transverse, classical,
telegraph noise. The qubit diagonal splitting is driven as a power law, 
$\Delta_\nu(t)=\alpha_\nu |\alpha_\nu t|^\nu \sign(t)$, with driving rate $\alpha_\nu$,
where particular attention has been on the role of $\nu$.

An expression, Eq.~\eqref{z_p-infty}, for the state after transition, $z_p(\infty)$,
has been derived in the limit of fast noise, $\gamma\gg\alpha_\nu$. 
From this expression we have found that there exists 
a critical $\nu_c=1/2$ such that the system looses all coherence 
when $\nu<\nu_c$, even for very weak noise, $v\ll\alpha_\nu$. 
When $\nu>1$  some coherence is retained and for weak noise 
the end state is actually fully coherent, $z_p(\infty)=1$.
The same results also applies for transverse coherence (phase coherence).

For linear driving and fast noise, $z_p(\infty)$ is independent of 
noise switching rate $\gamma$. However, this property holds only 
for $\nu=1$ and for $\nu\neq 1$, $z_p(\infty)$ depends on $\gamma$ in the following way:
increasing $\gamma$ decreases end state coherence when $\nu<1$ and increases 
end state coherence when $\nu>1$. 

We have also studied the limit of slow telegraph noise, $\gamma\lesssim\alpha_\nu$. 
A critical $\nu_c$ seems to exist in that case, but the 
value is less than for fast noise, i.e., $\nu_c<1/2$ and it depends on $\gamma$.
An interesting property of strong and slow noise is that 
it can drive the system to the other diabatic level. In terms of 
coherence, this means that the system is driven through the origin of 
the Bloch sphere, representing full decoherence. After that, coherence 
increase with time. Strong and slow noise also experiences 
a nontrivial dependency on $v$ and $\gamma$. E.g., increasing noise 
strength can in some cases also lead to increasing $|z(\infty)|$, representing  
increased coherence.

\acknowledgements
This work was supported financially by The Norwegian Research
Council, Grant No. 158518/431 (NANOMAT). The work of
YG was partly supported by the U. S. Department of Energy Office of Science
through contract No. DE-AC02-06CH11357.


\begin{thebibliography}{30}
\bibitem[{not()}]{note:lz-slow-weak-convergence}
\bibinfo{note}{In principle higher order terms in $v/\alpha_\nu$ can diverge,
  even if this term converges, so the convergence of this integral gives only a
  lower bound on $\nu_c$.}
\end{thebibliography}
\end{document}